\documentclass[journal, 5p,times,twocolumn]{elsarticle}

\usepackage{lineno,hyperref}
\usepackage{epsfig}
\usepackage{epstopdf}
\usepackage{amsmath}
\usepackage{graphicx}
\usepackage{multicol}
\modulolinenumbers[5]

\begin{document}

\begin{frontmatter}

\title{DNN-HMM based Speaker Adaptive  Emotion Recognition using Proposed Epoch  and MFCC Features}
\author[rvt]{Md. Shah Fahad}
\ead{shah.cse16@nitp.ac.in}
\author[focal]{Jainath Yadav}
\ead{jainath@cub.ac.in}
\author[rvt]{Gyadhar Pradhan}
\ead{gdp@nitp.ac.in}
\author[rvt]{Akshay Deepak}
\ead{akshayd@nitp.ac.in}
\address[rvt]{Department of Computer Science,
National Institute of Technology
Patna, India }
\address[focal]{Department of Computer Science,
Central University of South Bihar,
Patna, India}


\begin{abstract}
 Speech is produced when time varying vocal tract system is excited with time varying excitation source.
 Therefore, the information present in a speech such as message, emotion, language, speaker is due to the combined effect of both excitation source and vocal tract system. However, there is very less utilization of excitation source features to recognize emotion. In our earlier work, we have proposed a novel method to extract glottal closure instants (GCIs) known as epochs. In this paper, we have explored epoch features namely instantaneous pitch, phase and strength of epochs for discriminating emotions. We have combined the excitation source features and the  well known Male-frequency cepstral coefficient (MFCC) features to develop an  emotion recognition system with improved performance. DNN-HMM  speaker adaptive models have been developed using MFCC, epoch and combined features. IEMOCAP  emotional database has been used to evaluate the models. The average accuracy for emotion recognition system when using MFCC and epoch features separately is 59.25\% and 54.52\% respectively. The recognition performance improves to 64.2\% when MFCC and epoch features are combined. 
\end{abstract}


\end{frontmatter}


\section{Introduction}   
Automatic emotion recognition from speech signal has  fascinated the research community in the recent years due to its applicability in real-life. Human beings use a lot of emotions along with textual messages to convey the intended information.  Emotions  improve human computer interactions (HCI) system such as interactive movies \cite{nakatsu2000emotion}, story telling and E-tutoring applications \cite{ververidis2003state},  and,  retrieval and indexing of the video/audio files \cite{sagar2007characterisation}. Emotion recognition system assists to improve the quality of service of call attendants at call centers \cite{lee2005toward}. Automatic emotion detection could be helpful in the psychological treatment as used in  references [\cite{ooi2012early},\cite{low2011detection},\cite{yang2013detecting}]. It can also be useful in the case of surveillance systems \cite{clavel2008fear}.
Modern speech-based systems are designed largely using neutral speech. Here, the components of emotions can be used as an add-on to improve the accuracy in practical applications.

Excitation source features are not much exploited to recognize emotions. Observation from the literature reveals that the majority of the previous works used prosodic and system features  for emotion recognition using speech [\cite{wang2004investigation,nicholson2000emotion}]. The system features MFCCs, Linear Predictive Cepstral Coefficients (LPCCs) and their derivatives reflect the emotion specific information. Prosodic features such as fundamental frequency, duration, energy and intonation are also used for emotion recognition. Combinations of prosodic and system features are also widely used for emotion recognition. Reference \cite{ververidis2004automatic} uses supra-segmental features such as energy, F0, formant locations, energy, dynamics of F0 and formant contours for emotion classification. The statistical parameters of F0 like maximum, minimum, and median values, and the slopes of F0 contours have emotion specific information \cite{dellaert1996recognizing}. However, not much work has been done in using excitation source features for emotion recognition.

Reference \cite{wang2004investigation} combined 55 features (24 MFCCs, 25 prosodic and 6 formant frequencies) for recognizing six emotions. Prosodic and spectral features are combined in reference \cite{nicholson2000emotion} for  emotion classification. It is proven from literature that a combination of different complement features improve the accuracy of emotion recognition system. Most of the features are extracted from speech based on the assumption that the speech signal is stationary in the small speech segment. However, the speech features  -- either source features or system features -- vary rapidly in emotional speech because of the rapid changes in the vibration of the vocal cords. In reference \cite{krothapalli2013characterization}, emotion recognition model is developed using a combination of epoch and MFCC features. The proposed model used zero frequency filter (ZFF) method for extracting epoch features. The accuracy of epoch detection using ZFF  decreases for emotional speech because it requires a priory pitch period  to detect epoch location. However, the pitch period of emotional speech varies frequently in an utterance. The emotion recognition model (in reference \cite{krothapalli2013characterization}) was developed using auto-associative neural networks (AANN) and support vector machines (SVM) on IITKGP-SESC database.  

In our earlier work \cite{yadav2017epoch}, we proposed a robust method to detect epoch locations. In this paper, epoch features namely instantaneous pitch, phase and strength of excitation (SOE) are extracted. These features are explored for different emotions and combined with MFCCs for classifying four emotions. Using this method, a significant increase in the accuracy of emotion recognition model was observed. The average accuracy for emotion recognition system when using MFCC and epoch features separately is 59.25\% and 54.52\% respectively. This improves to 64.2\% when MFCC and epoch features are combined. 
 
The rest of the paper is organized as follows. Section \ref{a} contains the description of speech databases, Sec. \ref{b} describes detection of epoch features and  Sec. \ref{c} briefly discusses MFCC and development of emotion recognition models. The results are discussed in Sec. \ref{d}. Section \ref{e} concludes the paper. 

\section{Databases}{\label{a}}
Our proposed model has been evaluated on IEMOCAP (Interactive emotional dyadic motion capture database)\cite{busso2008iemocap} and IITKGP:SEHSC (Indian Institute of  Technology Kharagpur:  Simulated Emotion Hindi Speech Corpus) \cite{koolagudi2011iitkgp}.
IEMOCAP database is a multi-modal database which contains audio, video, text and gesture information of conversations arranged in dyadic sessions. The database is recorded with ten actors (five male and five female) in five sessions. In each session, there are conversations of two actors, one from each gender, on two subjects. The conversation of one session is approximately five minutes long. The contents of the database are recorded in both scripted and spontaneous scenarios. The total number of utterances in the database are 10,039, where 4,784 utterances are from
the spontaneous sessions and 5,225 are from the scripted
sessions. The average duration of an utterance is 4.5 seconds while the average word count per utterance is 11.4 words.  The duration of the database is about 12 hours. The database is labeled as per the two popular schemes: discrete categorical label (i.e, labeled as happy, anger, neutral and sad) and continuous dimensional label (i.e, valence, activation and dominance). We have only used the audio tracks and the corresponding discrete categorical labels for emotion recognition.

  In IITKGP-SESC, fifteen emotionally neutral Hindi text prompts were used for recording the emotion in multiple sessions to capture diversity. In each session, 15 sentences in eight basic emotions are uttered by each artist. Recording was done with the help of SHURE dynamic cardioid microphone C660N at 16 kHz sampling frequency. The Hindi emotional speech database has 10 speakers (five males and five females) and 15 sentences were recorded for eight emotions (Neutral, Happy, Angry, Sad, Disgust, Sarcastic, Surprise and Fear). There are a total of 12000 speech utterances (10 speakers x 15 sentences x 8 emotions x 10 sessions) in the Hindi emotional speech database. There are 1500 articulations for each emotions. The number of syllables and words in the sentences lie  in the range of 9-17 and 4-7 respectively.
  
  \section{Extraction of  Epoch features using Zero time Windowing method}{\label{b}}
   In our method, voiced regions are detected using the phase of zero frequency filtered  speech signal \cite{kumar2016voice}. After that, Zero Time Windowing (ZTW) method \cite{bayya2013spectro} is applied to get Hilbert envelope of the Numerator Group Delay (HNGD) spectra of each of the voiced segments. The amplitude of the sum of the three prominent peaks is obtained from each spectrum of the HNGD. The resulting output reproduces the instantaneous energy profile of the windowed signal. The spectral energy profile, obtained from HNGD spectrum, shows high energy at the epoch locations because of high SNR (signal to noise ratio) at these locations. Further, the spectral energy profile is normalized using mean smooth filter. The normalized  spectral energy profile is then convolved with a Gaussian filter to highlight the peaks. The positive peaks  -- selected after removing the spurious peaks -- are considered as epochs. Next, each of the above step is described in detail.
   
   \subsection{Voiced Activity Detection (VAD)}  
    Epochs  are present in the voiced regions due to vibration of the vocal cords. Hence, we first divide the speech into voiced and unvoiced regions based on its characteristics. In the present paper, voiced regions are detected \cite{kumar2016voice} using the phase of Zero Frequency Filtered Signal (ZFFS). The ZFFS of a speech utterance is obtained by using zero frequency resonator \cite{murty2008epoch}. The phase of a ZFFS is determined using the Hilbrert transformation. Further, the phase-signal is split into  frames of size 30 ms with frame shift of 5 ms and  each frame is convolved with Hanning-window. The amplitude spectrum of Hanning-windowed frame is computed. Thereafter, the sum of the first 10 harmonics is computed. The decision of voiced and non-voiced regions is taken based on the appropriate threshold of global maxima of the  sum of phase harmonics (SPH) because the global maxima of  the SPH of voiced regions is significantly higher than unvoiced regions. 
   \begin{figure*}
   \begin{center}
   \includegraphics[width=1.0\linewidth]{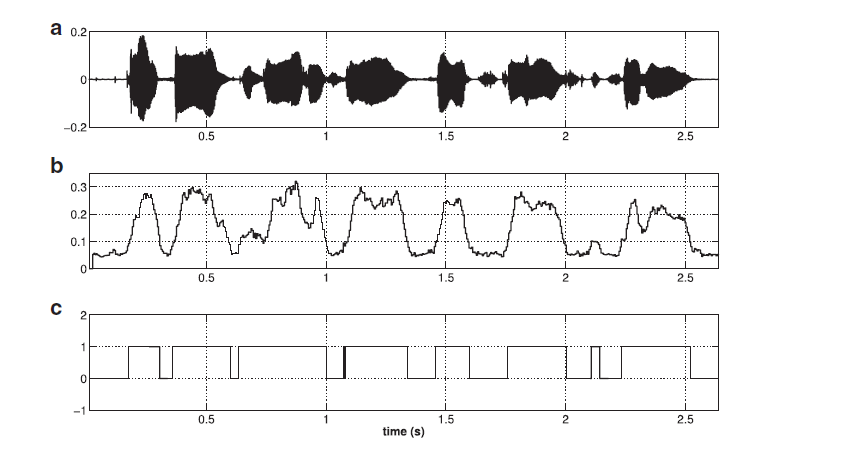}
   \caption{Detection of voiced and unvoiced regions using the phase of ZFFS. (a) Speech signal. (b) its corresponding global maxima of SPH. (c) unvoiced and voiced regions correspond to low and high amplitude respectively.}
   \label{m}
   \end{center}
   \end{figure*}
   
    Voiced and unvoiced regions of a speech signal are detected by setting the threshold of 0.08 for global maxima of SPH of each and every frame as shown in Fig. \ref{m}.
    Fig. \ref{m}(a) shows the the speech 
    signal. The corresponding global maxima of SPH is shown in
    Fig. \ref{m}(b), which is separated as voiced and unvoiced speech in Fig. \ref{m}(c) through rectangular waveform. Here, voiced speech is labeled 1(high) and unvoiced speech is labeled 0 (low).

    \subsection{Sequence of Steps for Epoch extraction }
      The steps to detect epoch locations are described next. 
      \begin{enumerate}
     
      \item The voiced segment is detected using  the phase of zero frequency filtered speech signal \cite{kumar2016voice}.
      
      \item The  voiced speech signal is differentiated to remove any low frequency bias in the speech signal using the formula 
      \begin{equation}
      y[n] = s[n]-s[n-1]
      \end{equation}
      where:\\
      $y[n]$ is the differentiated signal at $n^{\textit{th}}$ sample\\
      $s[n]$ is the actual speech signal at  $n^{\textit{th}}$ sample, and,\\
      $s[n-1]$ is the actual speech signal at  $(n-1)^{\textit{th}}$ sample\\
    
      \item Three milliseconds  segments  of the differentiated speech signal (resulting in $M$ = 48 samples) were taken at each sampling point. These were appended with $N-M$ (2048-48)  zeros  to  obtain sufficient resolution   in the frequency domain.
      
      \item The time domain signal is multiplied with the square of window function $h_{1}$ (defined below) to achieve the smoothened spectrum by integration in the frequency domain. 
          \begin{equation}
                 h_{1}[n] = \begin{cases} 0 & n = 0 \\  h_{1}[n] = \frac{1}{4 sin^{2}(\frac{\pi n}{N})}& n=1,2,..,N-1 \end{cases}
                     \end{equation}
      \item  The ripple effect due to truncation is reduced by multiplying the signal of the previous step  with the window $h_{2}$, which is defined as: 
        \begin{equation}
        h_{2}[n] = 4 cos^{2}(\frac{\pi n}{2M}), n = 0,1,2...,M-1
      \end{equation}
      The resultant signal $x[n]$ is called windowed signal. 
      \item  To highlight the spectral features, the numerator of group delay of windowed signal, denoted  $g[k]$,  is computed as:  
      \begin{equation}
        g[k] = X_{R}[k]Y_{R}[k]+X_{I}[k]Y_{R}[k], k= 0,1,2...,N-1   
        \end{equation}
       The resultant signal is known as DNGD signal.
      \item Hilbert envelope of the DNGD spectrum is computed to prominently highlight the spectral peaks.  The Hilbert envelope $h_{e}[k]$ of DNGD signal $g[k]$ is computed as: 
         
         \begin{equation}
        h_{e}[k] = \sqrt{g^{2}[k]+g_{h}^{2}[k]}
         \end{equation}
         
         where $g_{h}[k]$ is the Hilbert transformation of the sequence $g[k]$. It is computed as:
         
         \begin{equation}
              g_{h}[k] = IDFT{E_{h}(w)}
               \end{equation}
               where  $E(\omega)$ is the DTFT of the sequence $g(k)$. It  is defined as:
               \begin{equation}
              E_{h}(\omega) = \begin{cases} -jE(\omega), & 0 < \omega < \pi \\     jE(\omega, & -\pi < \omega <0 \end{cases}
               \end{equation}

      \item  The sum of the three most prominent peaks of the HNGD spectrum is determined at each sampling instant. The resultant amplitude  shows high SNR around glottal closure. Further, the amplitude contour is smoothened using 5-point mean smoothing filter to eliminate any outliers.

      \item   The sum of  the three prominent peaks obtained from each HNGD spectra is called spectral energy profile. The spectral energy profile is convolved with a Gaussian filter of size, average pitch period of that segment. A Gaussian filter of length L is given by
      \begin{equation}
      G[n] = \frac{1}{\sqrt{2\pi\sigma}}e^{-\frac{n^{2}}{2\sigma^{2}}}, n = 1,2,...,L
      \end{equation}
      The standard deviation $\sigma $ used in the above formula is $\frac{1}{4^{th}}$ of the Gaussian filter length.
      
      \item The spurious peaks are eliminated by using following sub steps:
     
       \begin{itemize}
       \item[(a)] First, the spurious peaks are eliminated on the basis that the difference between successive peaks should not be less than 2 ms. This is because 2ms is the minimum range of the pitch period. If two successive peaks having a difference of less than 2ms are found, the peak location with less amplitude is removed.
       
       \item[(b)] Two successive peaks bound a negative region between them. This criteria also eliminates some spurious peak locations.
       \end{itemize}
     
     \item The positive peaks in epoch evidence plot represent epoch locations.
     
       \end{enumerate}

\begin{figure}[h]
                        \begin{center}
                      \includegraphics[width=1.0\linewidth]{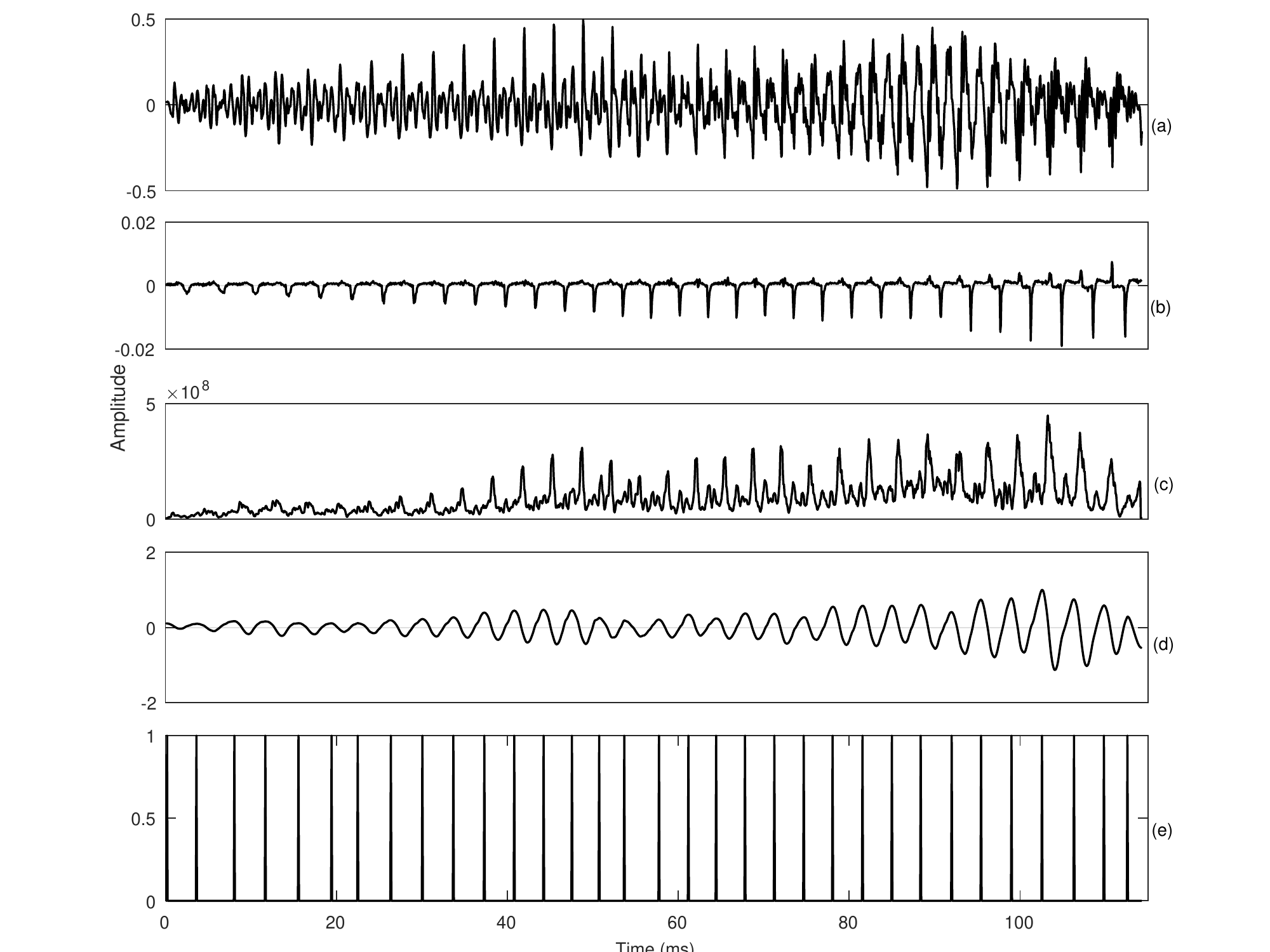}
                     \caption{Epoch extraction using proposed method. (a) Angry speech segment. (b) Differentiated EGG signal. (c) Spectral energy profile obtained from HNGD spectrum. (d) Epoch evidence plot. (e) Epoch locations.}
                     \label{j}
                         \end{center}
                         \end{figure}

    Epoch detection using ZTW method is shown in Fig. \ref{j}. The angry emotional speech segment is shown in Fig. \ref{j}(a) and its differentiated EGG signal is shown in Fig. \ref{j}(b). The spectral energy profile obtained from HNGD spectrum of the speech signal using ZTW analysis is plotted in Fig. \ref{j}(c). The epoch evidence plot  after convolving spectral energy profile with a Gaussian window of 2 m sec is shown in Fig. \ref{j}(d). Epoch  locations are shown in Fig. \ref{j}(e).
       
        ZTW method for epoch detection is robust for emotional speech \cite{yadav2017epoch}.  This method is based on spectral peak energy, therefore, it preserves the energy of the signal.

        \subsection{Epoch Features}
        The epoch features such as instantaneous pitch, strength of the epoch, slope of the strength of the epoch, the change of phase at the epoch are specific to each emotion \cite{krothapalli2013characterization}. The above mentioned features are determined by the epoch signal obtained by ZTW method \cite{yadav2017epoch}. The advantage of this method is that the value at epoch location is actually the sum of the glottal formants. Therefore, the epochs retain both time and spectral information.
        
        \subsubsection{Instantaneous Frequency}
         Instantaneous Period (IP) is the duration between two successive epoch locations; instantaneous frequency, denoted $\Delta f$, is computed as the reciprocal of IP \cite{koolagudi2010emotion,narendra2015robust}:
        \begin{equation}
        \Delta f = \frac{1}{t(i)-t(i+1)},i=1,2,....(n-1)
        \end{equation}
        where $t(i)$ represents $i^{th}$ epoch location.
        
        \subsubsection{Strength Of Excitation}
        The Strength Of Excitation (SOE) is computed as the difference between two successive epoch values \cite{gangamohan2014excitation}:
        \begin{equation}
        y(i) = {x(i)-x(i+1)},i=1,2,....(n-1)
        \end{equation}
        where $x(i)$ is the epoch strength at $i^{th}$ epoch.
        
        \subsubsection{Instantaneous Phase}
        The  instantaneous phase of a glottal signal is obtained by the cosine of the phase function of the corresponding analytical signal.
        \begin{itemize}
         \item The analytic signal $g_{a}(n)$ corresponding to glottal signal $g(n)$ is given by
        \begin{equation}
        g_{a}(n) = g(n) +jg_{h}(n)
        \end{equation}
        \item where $g_{h}(n)$ is the Hilbert transformation of $g(n)$, and is obtained by
        
             \begin{equation}
                  g_{h}[n] = IDFT{g_{h}(w)}
                   \end{equation}
                   where  ${g_{h}(w)}$ is defined as:
                   \begin{equation}
                  g_{h}(\omega) = \begin{cases} -jG(\omega), & 0 < \omega < \pi \\     jG(\omega, & -\pi < \omega <0 \end{cases}
                   \end{equation}
                   $G(\omega)$ is the DTFT of the sequence $g(n)$ and IDFT denotes Inverse Discrete Fourier Transform
                   and 
                   \item The Hilbert envelope of glottal signal $g(n)$ is calculated as:
                   \begin{equation}
                       h_{e}[n] = \sqrt{g^{2}[n]+g_{h}^{2}[n]}
                        \end{equation}
                  \item  The cosine of the phase of the analytic signal $g_{a}(n)$ is
                   given by
                   \begin{equation}
                   cos\Phi(n) = \frac{Re g_{a}(n)}{|g_{a}(n)|} =\frac{g(n)}{h_{e}[n]}
                   \end{equation}
                   where $g(n)$ is glottal signal derived from speech signal $s(n)$ using ZTW method.
                   \end{itemize}

\begin{figure*}[!ht]
        \begin{center}
         \includegraphics[width=1.0\linewidth]{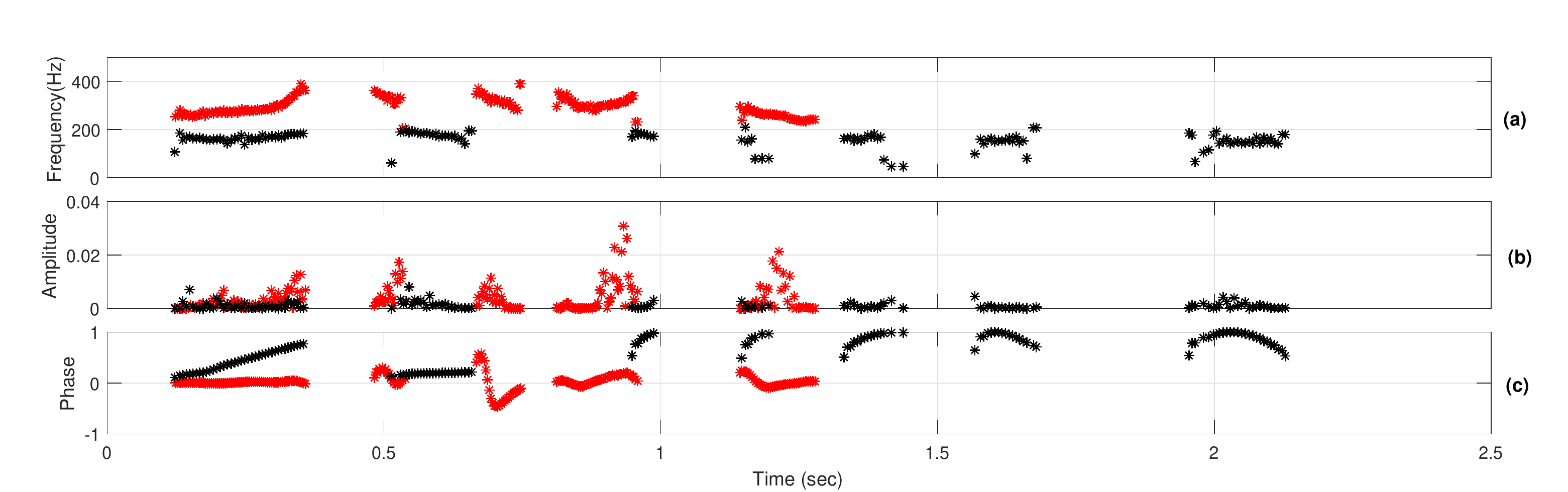}
        \caption{Instantaneous pitch and SOE contours of angry and sad speech signal using proposed method. (a) Instantaneous pitch contour, (b) SOE contour, and (c)  Instantaneous phase contour of angry and sad speech signal.}
        \label{k}
        \end{center}
        \end{figure*}

        In Fig. \ref{k}., instantaneous frequency  and SOE values of same speech utterance by same speaker in different emotions are plotted. Figure \ref{k}(a) shows instantaneous pitch for two emotions: angry and sad. Red color indicates angry emotion while black indicates Sad emotion. It is clear from Fig. \ref{k}(a) that the range of instantaneous pitch varies from 250-400 Hz for angry emotion while for sad it varies from 100-200 Hz. The instantaneous pith contour for same arousal emotion (happy and angry)  is same but their variation with time is different. This property of instantaneous pitch contour is well captured with dynamic model like Hidden Morkov Model (HMM) or Long Short Term Memory (LSTM) network.  Figure \ref{k}(b) shows SOE for two emotions: anger and sad. The variation of SOE is higher in angry emotion than sad emotion. The variation of SOE is quite less in the case of sad emotion. \ref{k}(b) shows the phase of glottal signal, it is high for sad compared than angry. The two features SOE and glottal phase also discriminate between same arousal emotion (happy and angry).
          
        \section{Development of Emotion Recognition System}
\begin{figure}
        	\begin{center}
        		\includegraphics[height= 14cm, width=8cm]{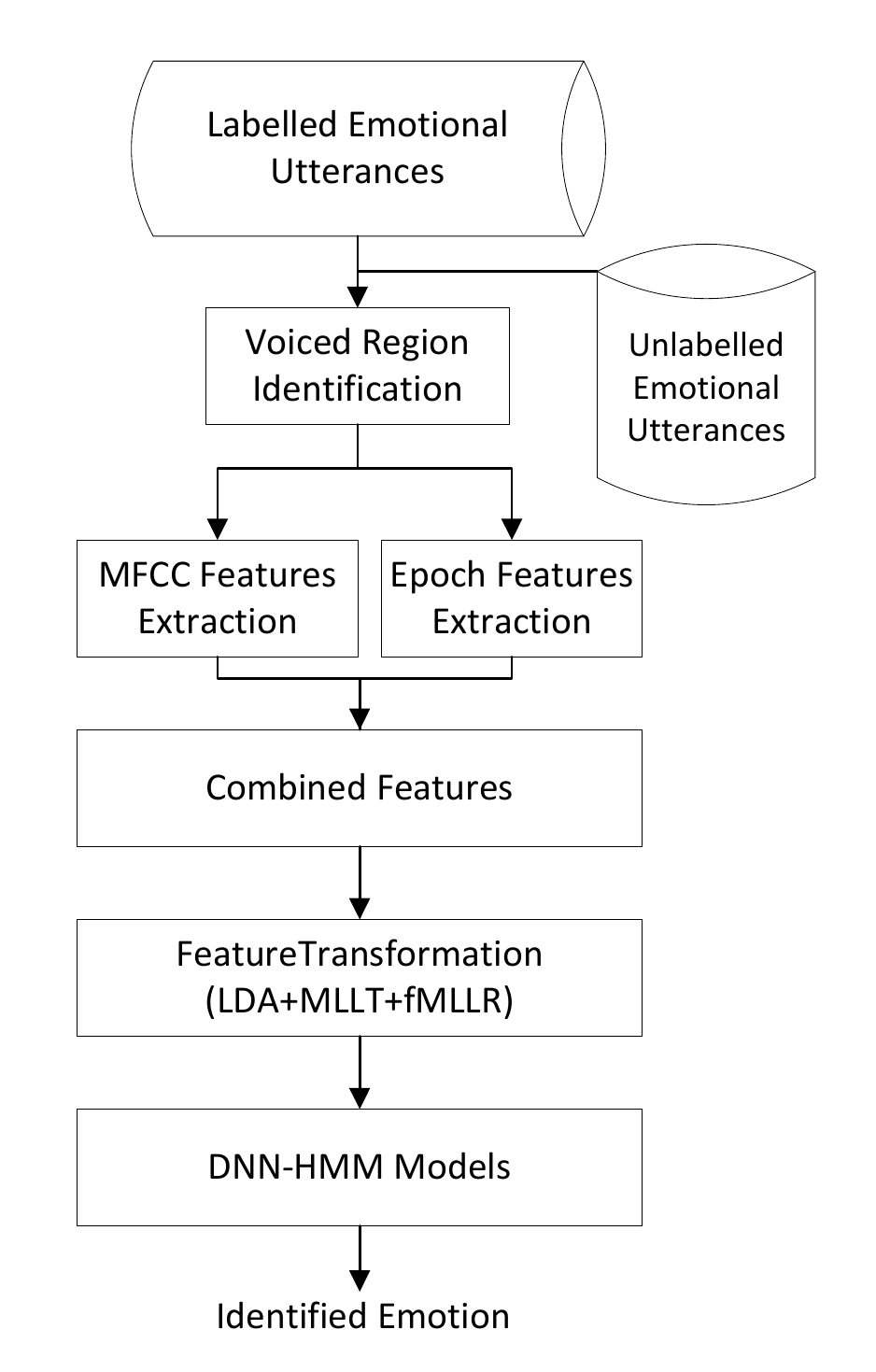}
        		\caption{Schematic diagram of the proposed emotion recognition model.  }
        		\label{l}
        	\end{center}
        \end{figure}

{\label{c}}
        Emotion recognition system is an outcome of two principal stages. In the first stage, training is performed using the features extracted form the known emotional speech utterances. In the second stage, i.e., the testing phase,  evaluation of the trained model is
        carried out on unseen emotional speech utterances. The schematic diagram of the proposed emotion recognition system is shown in Fig. \ref{l}. We combined the MFCC features with the epoch features namely instantaneous pitch, instantaneous phase and strength of epoch (SOE). The excitation source and system features have complementary information for recognizing emotions, hence, the combined features significantly improve the accuracy of emotion recognition.
        
        \subsection{MFCC Feature extraction}
\begin{figure*}
        \begin{center}
        \includegraphics[width=1.0\linewidth]{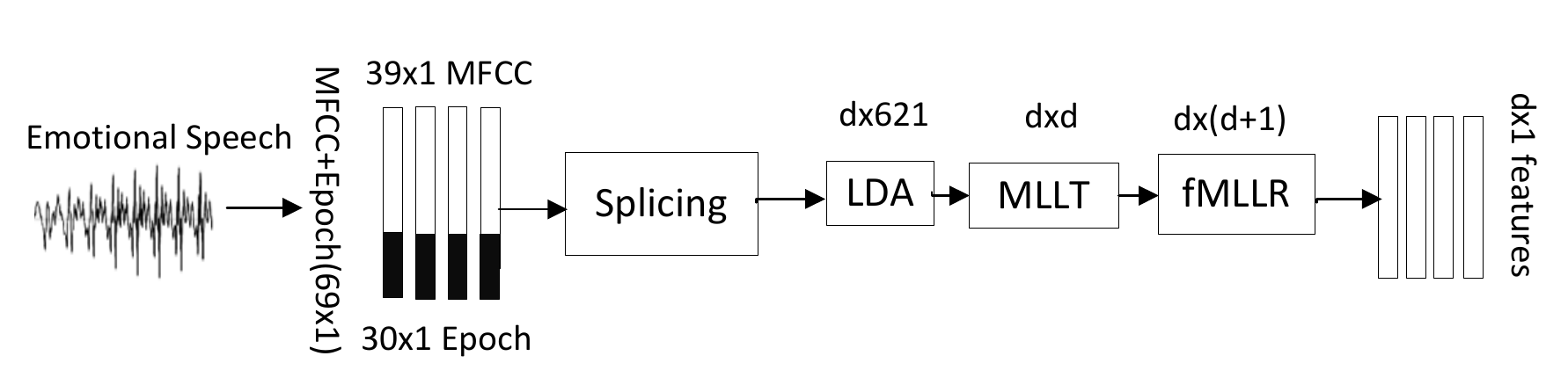}
        
        \caption{Schematic diagram of the proposed feature extraction and transformation.}
        \label{s}
        
        \end{center}
        \end{figure*}

        Mel Frequency Cepstral Coefficients (MFCCs) features also have emotion specific information. We combine MFCC features with epoch features in our model for recognizing emotions. Gradual spectral variations are captured using 13 MFCCs extracted from speech signal. The speech signal is segmented into frames of size 20 ms, where each frame is overlapped by 10 ms with the adjacent frame. For each frame, 13 MFCC features are extracted. To minimize spectral distortion at the beginning and at the end of each frame,  Hamming window is superimposed on each frame segment. MFCC features are extracted from these frames using the MFCC algorithm given in \cite{rabiner1993fundamentals}. Recording variations are countered by subtracting cepstral mean and normalizing variance of MFCCs at the utterance level. The schematic diagram of the proposed feature extraction and transformation is shown in Fig. \ref{s}. 
         
        \subsection{DNN-HMMs}
        In our work, the emotion recognition system has been developed using   Hidden morkov model (HMM) \cite{rabiner1986introduction} -- a dynamic modeling approach. It captures the temporal dynamic characteristics of different epoch features of corresponding emotions. In conventional HMM, the observation probabilities of HMM states are estimated by Gaussian mixture models (GMMs). The GMMs  used in such a conventional HMM are statistically inefficient to model non-linear data in the feature space. Therefore, we have replaced the GMMs with DNN to estimate the observation probabilities of observing input sequence at each state in the training phase. In this work, we have developed four HMMs for four discrete emotions.  Emotion label is assigned for an unknown speech utterance using Viterbi algorithm. The procedure for training and recognition of DNN-HMM is followed as mentioned in [\cite{li2013hybrid}, \cite{hinton2012deep}]. To the best of our knowledge, this is the first time that such a model is being used in an emotion recognition system. 
        
        For providing class labels to DNN, we used a GMM-HMM model with five states for each emotion class. Specifically, for each speech utterance in the raining set, viterbi algorithm is applied to find an optimal state sequence. The optimal state sequence is stored in the state-label mapping table, which is used to assign a label to each state. The training speech utterances, combined with their labeled state sequences, are then fed as input to the DNN. The output of the DNN is the posterior probabilities of the 20 output units. The observation probability of each state, denoted $p(i_{t}|q{t}) $, is calculated using Bayes theorem as follows:
        \begin{equation}
        p(i_{t}|q{t})  = \frac{p(q_{t}|i_{t})* p(i_{t}) }{p(q_{t})}
        \end{equation}
        Where $I = (i_{1},i_{2},.......i_{T} )$ is the input sequence and $p(q_{t}|i{t})$ is the posterior probability obtained as output from the DNN. During decoding, for an  unseen speech utterance, the probability of each emotion is estimated and the utterance is assigned the class whose estimated probability is maximum. $p(q_{t})$ is computed from the initial state level alignment  of the training set. $p(i_{t})$  remains constant because input feature vectors are assumed to be mutually independent.

        \section{Experimental Results and Discussion}{\label{d}}
        Three models were developed for emotion recognition: using system (MFFCCs) features, using source (epoch) features, and by combining MFCC and epoch features.  The model on combined features has significantly higher accuracy  compared to individual models.
        The experiments were performed on  IEMOCAP and  IITKGP:SEHSC databases. However, we have conducted experiments for only four emotions, namely angry, happy, sad and neutral. Three-fourth part of the database is used for  training purpose and the rest one-fourth of the database is used for evaluating the model. We have used MATLAB tool for feature extraction and  KALDI toolkit \cite{povey2011kaldi} for developing the system.  For the emotion recognition system developed using MFCC features, 13 MFCCs are extracted from each frame. Cepstral mean variance normalization (CMVN) \cite{viikki1998cepstral} is performed at utterance level to mitigate the recording variations.  We have also taken the derivative and double derivative of the normalized MFCCs as features. Therefore,  the total number of MFCC features for each frame is 39.  To preserve the contextual information, we have used the triphone model approach used in speech recognition where each frame is spliced with the left four frames and the right four frames. A significant improvement in emotion recognition accuracy is observed using the triphone model. Feature transformation is applied on the top of  9 spliced frame features. These features are projected into lower dimensional space using Linear Discriminant Analysis (LDA). Then, diagonalizing Maximum Likelihood Linear Transform (MLLT) [\cite{gales1998maximum,gales1999semi}] is applied to further improve the result. Speaker Adaptive Training (SAT) is also used to further
        enhance the accuracy of the emotion recognition model. For speaker adaptive training  Feature Space Maximum Likelihood Linear Regression (fMLLR) transformation is used during both  training and testing phases. Thus, accuracy of system is further improved using (LDA+MLLT+SAT)\cite{rath2013improved}. Four different DNN-HMM models corresponding to each emotion class are built using the transformed feature vectors.

        The DNN architecture used is: 80:512x5:20, where 80 is the number of transformed input features to the DNN and 512x5 represents 512 nodes in each of the 5 hidden layers. This DNN configuration was found to be optimal after experimenting with different sized configurations. The results discussed in this paper have been obtained on optimal DNN configuration only. There are 20 output classes in the DNN model (20=4x5, where 4 denotes the number of emotion classes and 5 denotes the number of states in HMM).  These output classes are treated as "ground-truth" states and are obtained by GMM-HMM based viterbi algorithm. The initial learning rate of 0.005 is gradually decreased to 0.0005 after 25 epochs. Additional 20 epochs are performed after this. The batch size for training is 512. The training of DNN is performed in three stages as in \cite{vesely2013sequence}: (i) unsupervised pre-training
        consisting in layer-wise training of Restricted Boltzmann Machines
        (RBM) by Contrastive Divergence algorithm; (ii) frame classification
        training based on mini-batch Stochastic Gradient
        Descent (SGD), optimizing frame cross-entropy; and (iii) sequence discriminative
        training consisting in SGD with per-sentence updates,
        optimizing state Minimum Bayes Risk (MBR).
        
        In our study, we have considered four categorical (class) labeled emotions namely angry, happy, sad and neutral. The numbers of utterances in each class are 1103, 595, 1084 and 1708 respectively with a total of 4490. The IEMOCAP database is imbalanced. 
         The model was trained in a speaker independent fashion. We used four sessions as training data and the remaining one session for testing. We followed the approach of leave-one-speaker-out cross-validation to generalize the model. The test dataset is also imbalanced corresponding to the emotion classes, hence, we calculated both weighted accuracy(WA)  and unweighted accuracy(UWA). Weighted accuracy is calculated by dividing  the total number of correct classified test examples with the total number of test samples. Unweighted accuracy is calculated for each emotion category and the average accuracy of all emotions class is taken. The unweighted accuracy is also called class accuracy.

        Similarly, for epcoh features, the emotion recognition system is developed using three epoch features namely instantaneous pitch, phase and the strength of epoch. These features  are extracted using ZTW method. We have taken frames of size 20 ms -- same as MFCC features -- to extract epoch features. The number of epoch features are different for each frame. To fix the length of epoch-feature vector, we have taken length as 10 -- the maximum number of epochs encountered in any frame. If the size of the feature vector is less than 10, we pad the remaining length with zeros. There are no adverse effects of padding to train the network because we transform the input feature vectors (using LDA+MLLT).  Therefore, the total number of epoch feature per frame is  30 (10 epochs $\times$ 3 features per epoch). We developed the DNN-HMM model for each emotion using these 30 epoch features. 
        
        Finally, we combined the epoch and MFCC features to improve the performance of emotion recognition system. After combining the MFCC and epoch features, the length of the feature vector becomes 69. 
         
          We have developed baseline GMM-HMM system using (1) monophone training, (2)  triphone training with $MFCC+\Delta+\Delta^{2}$, and (3)  triphone training with LDA+MLLT. We developed the DNN-HMM system with LDA+MLLT.  In Table \ref{o} we have shown the result of emotion recognition system using only MFCC and its derivative features. We have also applied LDA+MLLT transformation on MFCC and its derivative features. Our system is trained using both monophone and triphone training. Triphone system gives better result than monophone because it captures the contextual information. We also estimate the observation probability using DNN instead of GMM as described in  previous section. Our system gives best results in the case of DNN-HMM. The average accuracy increases approximately 3.5\% when observation probability of HMM models is calculated by DNN instead of GMM.    The confusion matrix for experiments done using only $MFCC+\Delta+\Delta^{2}$  features with LDA+MLLT transformation on DNN-HMM system is shown in Table \ref{f}. From the result it is clear that there is more confusion between angry and happy emotions because both are high arousal emotions. The sad and neutral emotions also show confusion because both are low arousal emotions. 
            
          \begin{table}[h]
          \centering
            \caption{Emotion classification performance (\%)  using the MFCC features on IEMOCAP database} \label{o}
            \label{h}
            \begin{tabular}{ccccc}
            \hline\vspace{1mm}
             \bf Features &\bf Model  \bf & \bf UWA ( \%)   \\\hline
           MFCC(monophone)  & GMM-HMM& 44.70 \\ \vspace{1mm}
           $MFCC+\Delta+\Delta^{2}$ (triphone)  & GMM-HMM& 47.70 & \\\vspace{1mm} 
          MFCC(LDA+MLLT)  & GMM-HMM & 51.25  \\ 
          MFCC(LDA+MLLT)  & DNN-HMM & 54.35  \\
            \hline             
           \end{tabular}
           \end{table} 
           \begin{table}[!h]
            \centering
            \caption{Emotion recognition performance on IEMOCAP Database, based on MFCC feature vector of voiced region using DNN-HMM. Abbreviations: A-Anger, H-Happy, N-Neutral, S-Sad}
            \label{f}
            \begin{tabular}{lllll}
              \hline\noalign{\smallskip}
               &\multicolumn{4}{l}{MFCC feature vector(Average: 59.58)} \\
              \cline{2-5}
               & A & H & N & S\\
               \hline
               \noalign{\smallskip}
              Anger   & \bf 60.21 & 23.29 & 9.45 & 7.05  \\ 
              Happy  & 26.56 &  \bf 58.17 &8.70 & 7.57  \\ 
              Neutral & 8.13 & 11.43 & \bf 59.71& 20.73  \\ 
              Sadness & 8.3 & 8.45 & 23.00 & \bf 60.25  \\ 
               \hline
            \end{tabular}
            \end{table} 
                  
           Similarly, we also developed the system for epoch features. The average recognition rate for the model developed using MFCC features only is 54.35\%.  The average recognition rate for the model developed using epoch features only is 54.15 \%. 
           
            \begin{table}
               \centering
               \caption{Emotion recognition performance on IEMOCAP Database, based on Epoch feature vector of voiced region. Abbreviations: A-Anger, H-Happy, N-Neutral, S-Sad}
               \label{t}
               \begin{tabular}{lllll}
                 \hline\noalign{\smallskip}
                  &\multicolumn{4}{l}{Epoch feature vector(Average: 54.52)} \\
                 \cline{2-5}
                  & A & H & N & S\\
                  \hline
                  \noalign{\smallskip}
                 Anger   & \bf 57.21 & 15.29 & 22.45 & 5.05  \\
                 Happy  & 13.56 &  \bf 52.24& 21.70 & 12.5  \\
                 Neutral & 15.23 & 14.40 & \bf 53.71& 16.66  \\
                 Sadness & 7.00 & 9.05 & 29.00 & \bf 54.95  \\
                  \hline
               \end{tabular}
               \end{table}
               
                The confusion matrix in Table \ref{t} shows the recognition performance for each emotions  using Epoch features.
               The diagonal elements of the confusion matrix shows the recognition performance for individual emotions using epoch features. From experimental result it is clear that epoch features discriminate well between angry and happy emotions compared to MFCC features. 
               The average recognition rate for the model developed using the combination of MFCC and epoch features  is 60.14\%.
                  The performance of the model for each emotion using MFCC features, epoch features and combination of MFCC and epoch features is compared in Table \ref{h}. The combined features significantly improves the accuracy of emotion recognition.
           \begin{table}[h]
           \centering
             \caption{Emotion classification performance (\%)  using the Epoch, MFCC and Combined(MFCC+Epoch) features on IEMOCAP database}
             \label{h}
             \begin{tabular}{p{3.7cm}cc}\hline
             \bf Features &\bf Model  \bf & \bf UWA ( \%)   \\ \hline 
            Epoch Features+LDA+ MLLT(triphone)   & GMM-HMM& 50.25 \\\vspace{1mm}
            Epoch Features+LDA+ MLLT(triphone)& DNN-HMM& 54.15  \\ \vspace{1mm}
            Epoch Features+MFCC+ $\Delta+\Delta^{2}$ (LDA+MLLT)& GMM-HMM & \bf57.25   \\\vspace{1mm}
            Epoch Features+MFCC+ $\Delta+\Delta^{2}$(LDA+MLLT) & DNN-HMM & \bf60.14\\       
             \hline             
            \end{tabular}
            \end{table}

        \subsection{ Speaker Adaptation}
\begin{figure*}
                      \centering
                       \includegraphics[width=0.9\linewidth]{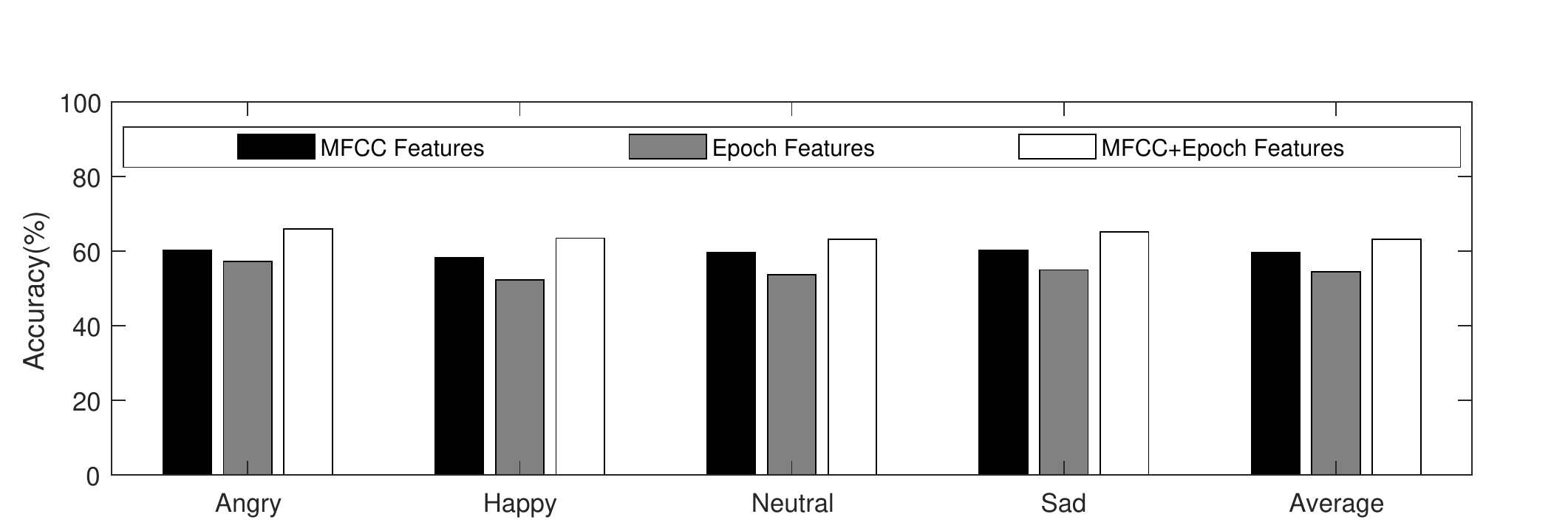}
                     \caption{Emotion classification performance (\%)  using the Epoch, MFCC and Combined(MFCC+Epoch) features on IEMOCAP database}\label{g}
                     \end{figure*}

        Adaptation is a necessary task  for emotion recognition. In general, we train our model with limited dataset but in real environment there may be different types of speakers and noise. There must be a robust method to adapt trained model in real environment. In this paper, we have applied Cepstral mean variance normalization (CMVN) at utterance level to mitigate the recording variations. fMLLR transformation is applied per speaker to adapt the emotion variation of different speakers. After the LDA-MLLT transformation of a feature vector, we transform this matrix into feature space using constraint maximum likelihood linear regression(CMLLR). The model is developed using the strategy of leave-one-speaker-out cross-validation where each time two speakers -- that were not a part of the training dataset -- are used for testing.
           \begin{table}[!hbt]
             \centering
               \caption{Emotion classification performance (\%)  using the Epoch, MFCC and Combined(MFCC+Epoch) features on IEMOCAP database}
               \label{u}
               \begin{tabular}{lcc}
               \hline
               \bf Features &\bf Model  \bf & \bf UWA(\%)   \\ \hline
               \vspace{1mm}
              MFCC(LDA+MLLT) & DNN-HMM & 54.35   \\\vspace{1mm}
              Epoch(LDA+MLLT)  & DNN-HMM & 54.15 \\  \vspace{1mm}    
              MFCC+Epoch(LDA+MLLT)  & DNN-HMM & 60.14  \\ \vspace{1mm}   
              MFCC(LDA+MLLT+SAT)  & DNN-HMM & \bf59.58  \\ \vspace{1mm}
              Epoch (LDA+MLLT+SAT)  & DNN-HMM & \bf54.52 \\\vspace{1mm}
              MFCC+Epoch(LDA+MLLT+SAT)&DNN-HMM & \bf64.20\\ \hline             
              \end{tabular}
              \end{table}
               There is significant improvement in recognition rate after applying speaker adaptive training for MFCC features. It is mentioned in the Table \ref{u} that after applying fMLLR the emotion recognition rate increases up to 4 \% for MFCC features but there is no improvement for epoch features. Therefore, we can say that epoch features are speaker independent features for which no speaker adaptive technique is required. The bar graph in Fig. \ref{g} Shows that emotion recognition accuracy is higher for combined (MFCC+Epoch) set of features than using each feature-set alone. 
               The average performance of combined features is increased by 5.34\% compared to the emotion recognition model  developed using MFCC features only. This result proves that both system features and excitation source features have complementary information for emotion recognition.

                     \begin{figure}
                     \centering
                  \includegraphics[width=1.0\linewidth]{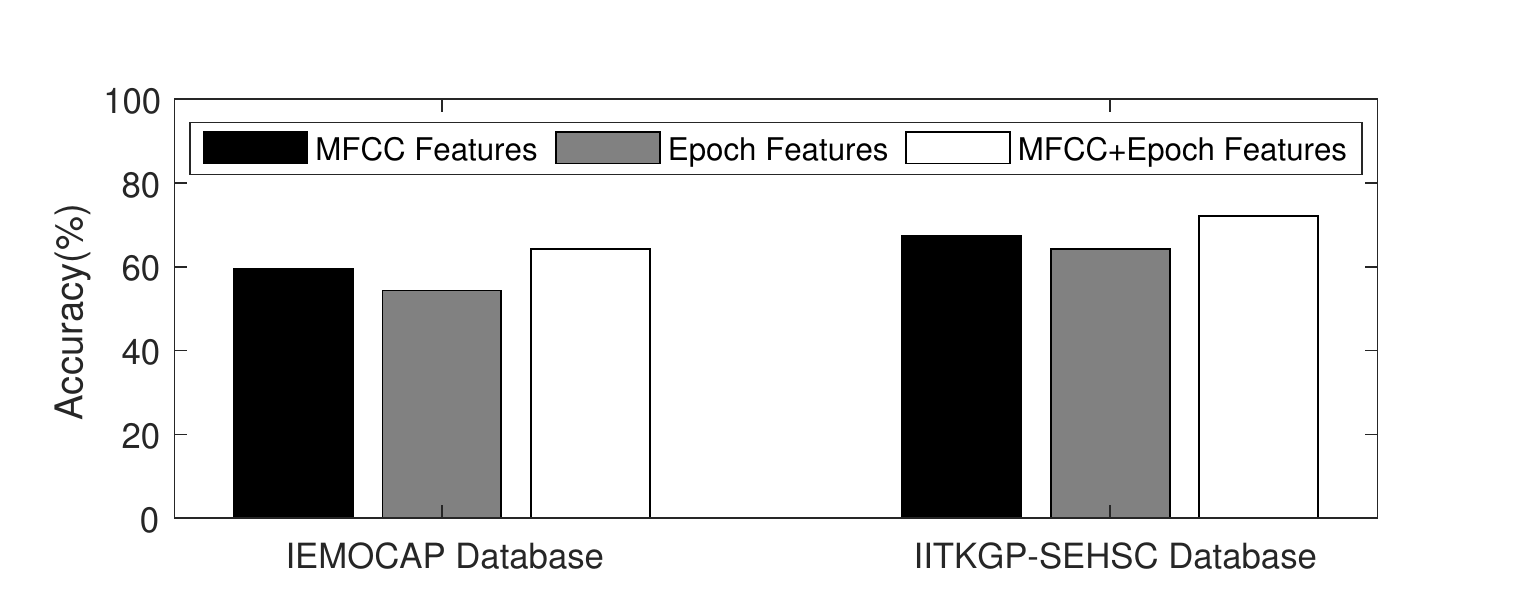}
                  \caption{Emotion classification performance (\%)  using the Epoch, MFCC and Combined(MFCC+Epoch) features on IEMOCAP and IITKGP-SEHSC databases}\label{i}
                   \end{figure} 
                    We evaluate our proposed approach on two databases IEMOCAP and IITKGP:SEHSC database. The bar graph corresponding to MFCC, Epoch and combined feature set for each database is shown in Fig.\ref{i}. In both databases, the accuracy increased for combined features. The accuracy is more in IITKGP:SEHSC database because it is a scripted database. IEMOCAP database contains both scripted and spontaneous sessions, and is more natural. It is also text independent database. The utterance length in IITKGP:SEHSC database is almost equal whereas large variation in utterance length is observed in IEMOCAP database.

                    We compare our result with the prior work on IEMOCAP database. In \cite{han2014speech}, DNN was used to extract the features from speech segment and further utterance level features were constructed and fed to the Extreme Learning Machine(ELM). In \cite{mirsamadi2017automatic}, raw spectrogram and Low Level Descriptors(LLDs) features were modeled with attentive LSTM. The accuracy is more for LLDs compared to spectrogram. In \cite{fayek2017evaluating} Convolutional Neural network(CNN)  was used for  feature extraction from speech frame and these features was fed to the dense neural network. \cite{satt2017efficient}, Long Short Term Memory (LSTM) network was used to preserve the contextual information of CNN-based features.  CNN-based features are fully data driven. It extracts  features from the raw spectrogram which are the representation of speech but it does not contain temporal resolution properly. To achieve temporal resolution, we have to restrict frequency resolution. All the methods used spectrogram for speech representation but it can mislead the accuracy. Our feature extraction approach is not data driven, we are identifying desired temporal and spectral features using signal processing technique. The HMM model captures the contextual information of epoch features.  As can be seen from the Table \ref{q} both the weighted and unweighted accuracy outperform from the other methods. Our result proves that MFCC and source features(epoch) contain complimentary information. 
                     \begin{table*}[t]
                              \centering
                                \caption{ Comparision of Emotion classification performance (\%)  reported in the prior work on IEMOCAP database} \label{q}
                                \begin{tabular}{cp{2.7cm}cc}
                                \hline\vspace{1mm}
                                 \bf Model &\bf Features &\bf WA(\%)  &\bf UWA ( \%)   \\\hline
                                  \vspace{1mm}
                              
                               DNN+ELM \cite{han2014speech}  & MFCC features, pitch-based features and their derivatives& 54.3 & 48.00 \\\vspace{1mm}
                               LSTM with attention \cite{mirsamadi2017automatic} & Local level descriptor and Spectrogram & 63.5 & 58.8 \\ \vspace{1mm} 
                              CNN \cite{fayek2017evaluating} & Spectrogram & 64.78  & 60.89 \\ \vspace{1mm}
                               CNN+LSTM \cite{satt2017efficient}  & frame-level Spectrogram & 68.8 & 59.4  \\
                               DNN-HMM  & Epoch (Proposed)  & 58.60 & 54.52  \\
                               DNN-HMM  & MFCC  (proposed) & 64.3 & 59.58  \\
                               
                               DNN-HMM  & MFCC+Epoch (Proposed)  & 69.5 & 64.2  \\
                               
                                \hline             
                               \end{tabular}
                               \end{table*}

\section{SUMMARY AND CONCLUSION}{\label{e}}
The paper highlights the robust characteristic of ZTW method for extracting epoch features. The DNN-HMM model is developed for each emotion using epoch features such as instantaneous pitch, strength of epoch (SOE). The average emotion recognition rate of the proposed model using epoch features is 54.52\%. The model developed using epoch features is further combined with the model developed using MFCC feature vectors. The observed accuracy of the proposed model using MFCC and epoch features together is 64.20\%. The experimental results show that the epoch feature set is complementary to the MFCC feature set for emotion classification. Our future work is to use LSTM network to capture the contextual information of epoch feature and to explore epoch features in the other applications
of speech processing such as speaker identification, speech recognition and, synthesis and language identification.

\section*{References}

\bibliography{acnref}

\end{document}